\documentstyle[aps,prl,twocolumn,epsfig]{revtex}
\begin{document} 
\draft
\twocolumn[\hsize\textwidth\columnwidth\hsize\csname @twocolumnfalse\endcsname
\title{
Geometrically Induced Multiple Coulomb Blockade Gaps
}
\author{
Mincheol Shin,\cite{e-mail} Seongjae Lee, Kyoung Wan Park, and El-Hang Lee
}
\address{
Electronics and Telecommunications Research Institute\\
Yusong POB 106, Taejon 305-600, Republic of Korea
}
\date{\today}
\maketitle

\begin{abstract}
We have theoretically investigated the transport properties of a
ring-shaped array of small tunnel junctions, which is weakly coupled
to the drain electrode. We have found that the long range interaction
together with the semi-isolation of the array bring about the
formation of stable standing configurations of electrons. The stable
configurations break up during each transition from odd to even number
of trapped electrons, leading to multiple Coulomb blockade gaps in the
the $I-V$ characteristics of the system.
\end{abstract}
\pacs{73.23.Hk, 73.23.-b}
]
\narrowtext 

%
% Introduction
%

\begin{figure}
\begin{center}
\leavevmode
\epsfig{width=0.95\linewidth,file=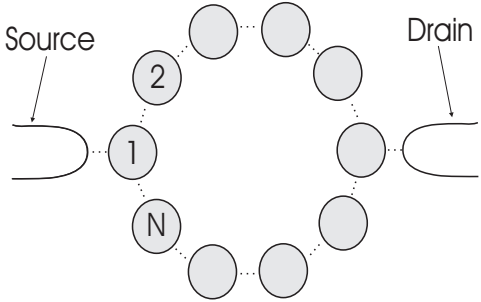}
\caption{The two-dimensional array of size $N$. Dashed lines represent
tunneling junctions between neighboring dots.}
\label{fig:loop}
\end{center}
\end{figure}

Transport properties of arrays of small tunnel junctions, such as the
Coulomb blockade effect and correlated single electron tunneling, have
been studied
extensively.\cite{general,Amman,Hu,Bakhvalov,Whan,Likharev_Matsuoka}
Most of the studies, however, were carried out for one-dimensional (1D)
arrays and were based on the nearest neighbor interaction
approximation.\cite{general,Amman,Hu,Bakhvalov} The soliton potential
in that case is known to decay exponentially with the screening length
$\sim \sqrt{C/C_0}$ where $C$ and $C_0$ are the junction capacitance
between neighboring dots and the self-capacitance of a dot,
respectively. According to recent studies, however, the soliton
potential in one- or two-dimensional arrays should decay as $1/r$,
where $r$ is the distance, if the full interaction between dots in the
array is taken into account.\cite{Whan,Likharev_Matsuoka} In this
work, we have applied the full interaction result to a ring-shaped
array as shown in Fig.\ \ref{fig:loop}, which has two branches or
paths for electrons between the source and the drain electrodes. Due
to the long range interaction between dots, the charge distribution of
one branch of the array is expected to substantially affect that of
the other branch, which should considerably influence the transport
property of the array.

Furthermore, we have isolated the array from the drain to some degree in
order to trap the electrons more easily within the array. These two
ingredients, i.e., the long range interaction and the semi-isolation of the
array, will be shown to lead to multiple Coulomb blockade gaps in
$I-V$ characteristics under some conditions, which will the main topic
in this paper.

%
% Theory
%

In the semiclassical approach that we have adopted
here,\cite{general,Amman,Hu,Bakhvalov} the charge $Q_i$ of the $i$-th
dot of the system of Fig.\ \ref{fig:loop}, where nearest neighboring
dots are coupled by a tunneling junction, is given by
\begin{equation}
\label{eq:Q_i}
\sum_{j} A_{ij}\phi_j= Q_i,
\end{equation}
where $\phi_i$ is the potential of the $i$-th dot and where
\begin{eqnarray}
A_{ij} &=& -C_{ij}, \ \ \ \ \ \ \ i \neq j \\
A_{ii} &=& \sum_{j} C_{ij}
\end{eqnarray}
where $C_{ij}$ is the capacitance between the $i$-th and $j$-th dots. We
have modeled the $C_{ij}$ as follows:
\begin{equation}
\label{eq:C_ij}
C_{ij} = \cases{C_0 \ \ \ \ \ \ \rm{for} \ \ \it{i} = j,\cr
                C \ \ \ \ \ \ \rm{for} \ \ \it{r_{ij}} = a,\cr
                C' a/(r_{ij} - a) \ \ \ \ \ \ \rm{otherwise}, \cr}
\end{equation}
where $a$ is the distance between the neighboring dots (assumed to be
the same for all neighbors) and $r_{ij}$ is the real, two-dimensional
distance between the $i$-th and $j$-th dots. The soliton potential in
this full-interaction model decays as $1/r$, where $r$ is the distance
between dots, so as to be consistent with the recent full
interaction studies.\cite{Whan,Likharev_Matsuoka} The parameter
$C'$ determines the degree of screening ability of nearby
dots. The range $0.1 \lesssim C'/C \lesssim 0.3$ seems to be reasonable
for semiconductor arrays with poor screening.\cite{Whan}

To isolate the array from the drain to some degree, the capacitance
between the drain electrode and each dot in the array is further
multiplied by a uniform factor $C_d/C$. We have set $C_d/C = 0.1$ in
this work, at which value the array is still conducting but coupled to
the drain sufficiently weakly.

The current $I$ is calculated for constant bias voltage $V$
between the source and the drain, via standard Monte Carlo
simulation.\cite{Amman,Bakhvalov} The transition rates are determined
by using the Golden-rule
formula, with the electrostatic energy of the form
\begin{equation}
E = {C_0 \over 2}\sum_{i}\phi_i^2+{1 \over
4}\sum_{i,j}C_{ij}(\phi_i-\phi_j)^2.
\end{equation}
In this paper, the unit of the current, the voltage, and the temperature
are $\bar{I} = e/RC, \bar{V} = e/C$, and $\bar{T} = e^2/k_B C$, respectively,
where $R$ is the inverse of the transmittance across the junction
between neighboring dots.

%
% Results
%

% Data 

\begin{figure}
\begin{center}
\leavevmode
\epsfig{width=0.8\linewidth,file=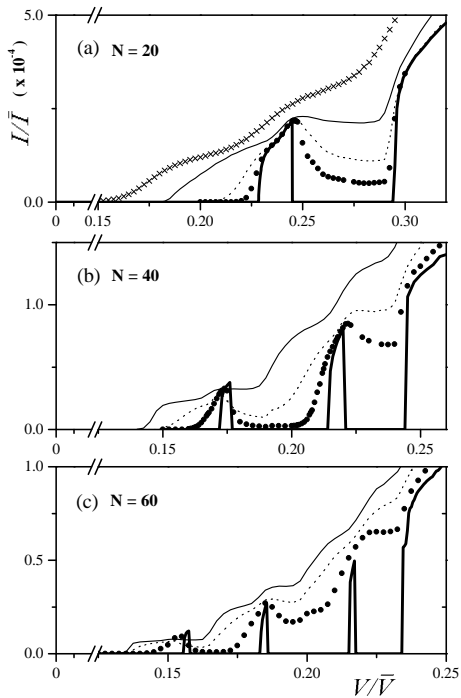}
\caption{The $I-V$ characteristics of the 2D array with 20 (a), 40 (b),
and 60 (c) dots at various temperatures. $C'/C = 0.25$ for all, but
$C_0/C = 0.6$ for $N=20$ and 0.7 for $N=40$ and $60$. The temperature
$T/\bar{T}$ = 0.005 and 0.0005 for crosses of (a) and for thin full line
of (c), respectively. Otherwise, $T/\bar{T}$ = 0 (thick full line),
0.0001 (solid circles), 0.00025 (dots), 0.001 (thin full line).
}
\label{fig:IV}
\end{center}
\end{figure}

We show in Fig.\ \ref{fig:IV} the $I-V$ characteristics of the
ring-shaped array with $N$ = 20 (a), 40 (b), and 60 (c), at various
temperatures. In contrast to the simply-connected 1D array case where
there is only one Coulomb blockade gap in a low-bias voltage range,
the zero-temperature $I-V$ characteristics in Fig.\ \ref{fig:IV}
exhibit sharp {\it multiple gaps}: there are two gaps for the array
with 20-dots, three for 40-dots, and four for 60-dots. These gaps at
zero temperature are transformed into negative differential
conductance (NDC) regions at finite temperatures, ultimately showing
monotonically increasing behavior at higher temperatures.

% <n> and I

\begin{figure}
\begin{center}
\leavevmode
\epsfig{width=1.0\linewidth,file=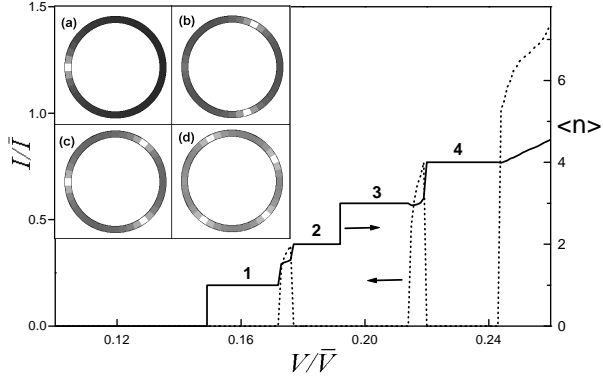}
\caption{The average number of electrons $\langle n \rangle$ (right
axis) and corresponding $I/\bar I$ (left axis) versus bias voltage for
the 40-dot array of Fig.\ \protect{\ref{fig:IV}}-(b) at zero
temperature. The potential profiles at the plateaus of $\langle n
\rangle$ = 1 (a), 2 (b), 3 (c), and 4 (d) are also shown, with bright
spots representing trapped electrons in the array.}
\label{fig:IVN}
\end{center}
\end{figure}

In Fig.\ \ref{fig:IVN}, we have shown $I/\bar I$ and the average
number of electrons $\langle n \rangle$ within the 40-dot array at
zero temperature. $\langle n \rangle$ shows clear steps with integer
multiples up to $\langle n \rangle = 4$, beyond which it increases
monotonically without showing further steps, as does the
current. The potential profiles at the plateaus of $\langle n \rangle$
= 1, 2, 3, and 4 for the 40-dot array are shown in Fig.\
\ref{fig:IVN}-(a), (b), (c), and (d), respectively. From the
figure, it is clear that the current peaks arise during the transient
phase in which $\langle n \rangle$ changes from odd to even ($1
\rightarrow 2$ and $3 \rightarrow 4$ for the first and second peaks)
and the multiple Coulomb gaps are separated by those peaks. Likewise,
the peak for the 20-dot array arises during the $1 \rightarrow 2$
transition, and the peaks for the 60-dot show up during $1\rightarrow
2$, $3\rightarrow 4$, and $5\rightarrow 6$ transitions,
respectively. The corresponding potential profiles at the plateaus are
similar to the ones in Fig.\ \ref{fig:IVN}.

% Even -> Odd and Odd -> Even Transitions

The charge configurations shown in Fig.\ \ref{fig:IVN} for the
40-dot array are stable insulating configurations with integer number
of trapped electrons within the array. As the bias voltage increases
from zero, the system undergoes successive transitions from one stable
configuration to another. If a transition occurs from even $\langle n
\rangle$ to odd $\langle n \rangle$, for instance, from $\langle n
\rangle = 2$ to $\langle n \rangle = 3$, the transition is smooth and
immediate: a new electron which just tunneled into the array to make
$\langle n \rangle = 3$ merely pushes the already built-up standing
charge configuration of two electrons a bit toward the drain and the
newly tunneled electron stays on dot 1, making another stable
configuration of three electrons (see Fig.\ \ref{fig:IVN}-(b) and
(c)). However, when a transition occurs from odd $\langle n \rangle$ to
even $\langle n \rangle$, for instance from $\langle n \rangle = 1$ to
$\langle n \rangle = 2$, the transition takes place via intermediate
{\it unstable} states. As the bias voltage increases such that tunneling
of the second electron into dot 1 from the source is inevitable,
the electron which has been on dot 1 is pushed and moves toward the
drain. Then possible charge configurations are: either one electron
solely travels toward the drain before another tunnels into the array,
or one electron in one branch travels ahead of the other electron in
the other branch. While these unbalanced charge configurations
persist, the system becomes conducting, until the stable configuration
of two electrons as shown in Fig.\ \ref{fig:IVN}-(b) is
eventually established at a higher bias voltage. That is why the
current peaks show up during the transient phase from odd to even
number of $\langle n \rangle$, whereas no current peaks arise during
the transition from even to odd $\langle n \rangle$.

% Analogy to Melting of a Solid

We may draw an analogy for the behavior of the ring-type array to
melting of a solid. The state when the standing configuration is built
up may be compared to a rigid solid state with symmetry. When the
externally-driven distortion of the lattice exceeds some threshold,
the symmetry is broken and the solid starts to melt. Likewise, the
bias voltage in our array drives the system toward broken-symmetry
state (the transient state during which $\langle n \rangle$ changes
from odd to even) and the array becomes conducting. In this case, the
symmetry is imposed by the special geometry which has two
branches. One difference in this analogy is that the array system
becomes insulating again when the symmetry of the system is restored
with higher integer value of $\langle n \rangle$.

% Effect of Thermal Fluctuation

The instability of the transient region (from odd to even $\langle n
\rangle$) is increased by the thermal fluctuations, so its width
broadens as the temperature is raised. Therefore, the current peaks,
whose widths depend on that of corresponding transient regions, also
broaden at higher temperatures, as can be seen in Fig.\
\ref{fig:IV}. The thermal fluctuations also destabilize the standing
configurations themselves established at zero temperature. The thermal
fluctuation gives the trapped electrons making up the stable
configurations a small but finite probability to tunnel through nearby
junctions. Once such tunneling event takes place, the balanced
configuration is temporarily broken and a trapped electron exits
through the drain, and the non-zero current flows in the additional
Coulomb blockade region, resulting in NDC behavior in the $I-V$
curve.

% Stability

We can estimate the degree of stability of a charge configuration by
calculating the free energy changes $\Delta F$ for its transitions to
adjacent configurations. For each of the stable configurations
discussed above, $\Delta F > 0$ for all possible transitions to
adjacent configurations: that is, they are local minima of the free
energy in the configuration space. To illustrate this point, let us
consider the simplest possible case of the 4-dot array with identical
junctions, where the stable configuration responsible for the second
Coulomb gap is $\{0,-1,0,-1\}$ (i.e., when electrons are at the second
and the fourth dots - see Fig.\ \ref{fig:loop} for the dot
indices). If we calculate the free energy change $\Delta F$ for
transitions to adjacent configurations $\{0,0,-1,-1\}$ and
$\{0,-1,-1,0\}$, we have (for the case of strong screening, for
simplicity)
\begin{eqnarray}
\label{eq:delF}
\nonumber \Delta F(V) &=& \{C^2(C_0^2+C_0 C-4C^2)/2D\}e^2/C \\ &-&
\{2C^2(C_0+2C)(C_0+C)/2D\}eV,
\end{eqnarray}
where the first term represents the electrostatic energy change, the
second the work done by the voltage source, and $D \equiv (C_0^2+5C_0
C+6C^2)(C_0^2+5C_0 C+2C^2)$. Eq.\ (\ref{eq:delF}) shows that if $C_0 >
\tilde{C}_0 = (\sqrt{17}-1)C/2$ and if $V < \tilde{V} = e(C_0^2+C_0 C
- 4C^2)/2C(C_0+2C)(C_0+C)$, $\Delta F(V) > 0$. Thus, the configuration
with electrons at dots 2 and 4 is {\it locally stable}, unless
thermal fluctuations which overcome the free energy difference are
introduced, that is, if $T < T_c(V) = \Delta F(V)/k_B$. Further
analysis with free energy changes can also give the widths of the
multiple Coulomb gaps and peaks. For the 4-dot case, the
second Coulomb gap appears in the interval of ($V_t^{(2)},{\rm
min}(\tilde{V},V_t^{(3)})$), where $V_t^{(2)}$ and $V_t^{(3)}$ are the
threshold voltages for entrance of the second and the third electron
into the array, respectively, which are given by $V_t^{(2)} \approx
1/2C_0$ and $V_t^{(3)} \approx V_t^{(2)} + (C/C_{eff})^2e/C$, where
$C_{eff} = \sqrt{C_0^2+4C_0 C}$. Further details will be published
elsewhere.

% Effect of C_0/C at T=0

\begin{figure}
\begin{center}
\leavevmode
\epsfig{width=0.9\linewidth,file=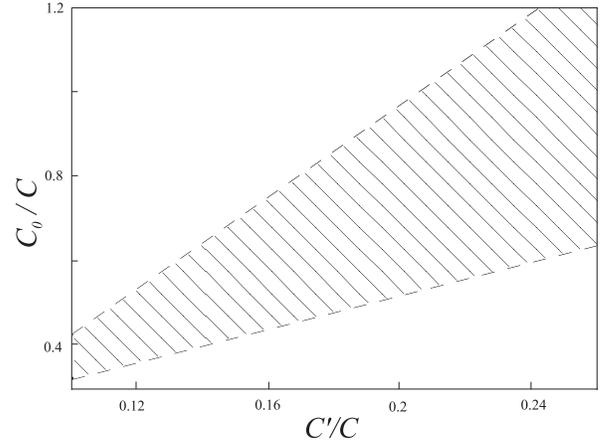}
\caption{The shaded region in the plot represents the region where the
multiple Coulomb gaps appear for the 20-dot array.}
\label{fig:CB_Region}
\end{center}
\end{figure}

As can be seen in the above analysis, $C_0/C$ is an important factor
for existence of the multiple Coulomb gaps in the $I-V$
characteristics. To observe the multiple Coulomb gaps separated by
peaks, $C_0/C$ should not be too weak nor too strong: in the
intermediate range of $0.1 \lesssim C_0/C \lesssim 1$ the multiple
Coulomb gaps are seen in general. The range is, however, dependent on
another factor $C'/C$ which reflects the degree of screening by nearby
dots. In Fig.\ \ref{fig:CB_Region}, we have schematically drawn the
region, in the plane of $C_0/C$ and $C'/C$, where the multiple Coulomb
gaps appear for the 20-dot array at zero temperature. As one may
anticipate, the figure shows that with less screening, which implies
stronger interaction among dots, the shaded region becomes wider as it
shifts toward weaker coupling between dots. At finite temperatures,
the shaded region expands considerably because the thermal
fluctuations can give rise to appearance of peaks which were too
narrow to be seen at zero temperature.

% Peak Height

The heights of the current peaks are multiples of the first peak
height. In Fig.\ \ref{fig:IV}-(b), the ratio of the first and the
second peak heights for the 40-dot array is very close to 1:2. That is
because the current peaks at zero temperature reach their maxima just
before the standing configurations composed of even number of
electrons are built up, as Fig.\ \ref{fig:IVN} clearly
shows. Likewise, for the 60-dot array, the ratio of the current peak
heights are close to 1:2:3.

% Nature of Interaction

The multiple Coulomb gaps are observed only when the standing
configurations through interaction between equal number of electrons
in upper and lower branches of the array are built up. For arrays of
large size ($N \gtrsim 20$), the standing configurations as shown in
Fig.\ \ref{fig:IVN} are possible due to the long-range nature of
the interaction (Coulomb repulsion). For arrays of smaller size ($N
\lesssim 20$), even nearest neighbor interaction with exponentially
decaying soliton potential is sufficient to bring about the standing
configurations. However, due to the small array size, only one
additional gap is seen in this case and, in contrast to the case of
the 20-dot array of Fig.\ \ref{fig:IV}-(a) where the second gap appears when
$\langle n \rangle = 2$, the second gap in arrays with the nearest
neighbor approximation appears with higher $\langle n \rangle$, which
means that, due to the short-range interaction, more electrons are
needed to build up the standing configuration. Likewise, if the
condition of weak coupling of the ring-type array to the drain is
lifted (i.e., $C_d/C$ = 1), it becomes harder for the electrons to be
trapped inside the array such that, as in the case of nearest neighbor
interaction only, the standing configurations appear with $\langle n
\rangle$ higher than in arrays with weak coupling to the
drain.

% Effect of Imperfection

The multiple Coulomb gaps are quite robust against possible
imperfection of the array. In real experiments, one can hardly expect
the dots in the array to be identical nor the array itself to be
perfectly symmetrical. The effect of those imperfections may be
reflected in our simulation simply by allowing the mutual capacitances
given by Eq.\ (\ref{eq:C_ij}) to have `random' contributions to some
degree. That is, $C_{ij} \rightarrow C_{ij}(1+\alpha \zeta_{ij})$,
where $\alpha$ is a constant adjusting the magnitude of the randomness
and $\zeta$'s are random numbers between -1/2 and 1/2. Note that we
have specifically put the subscripts for $\zeta$ to note that the
random numbers are differently assigned for different dots and
different pairs of dots. We have observed that for up to $\alpha =
0.1$ for the 20-dot array, the multiple Coulomb gaps are still seen (with
peak positions a bit shifted) or, at least, NDC regions are seen in
place of the multiple Coulomb gaps. If the source and the drain
electrodes are asymmetrically attached to the array, the multiple
Coulomb gaps or the NDC regions are seen as well. The stable charge
configurations when those disorders are introduced are similar to the
ones in Fig.\ \ref{fig:IVN} but with trapped electrons residing on
geometrically asymmetrical points as a consequence of their adjustment
to changes in electrostatic forces due to imperfections in the
array. That the multiple Coulomb gaps are robust against such
perturbations implies that details in modeling of the mutual
capacitance matrix are not important as long as the long range Coulomb
forces between dots are included.

% Conclusion

We have also investigated the possibility of observing the
multiple-gap behavior in arrays with two branches whose geometrical
shape departs from the circular one that we have considered here. Our
tentative conclusion is that the multiple Coulomb blockade gaps and
NDC behaviors seem to be generic features of arrays with two branches,
although the systematic behavior of $I-V$ with change of $\langle n
\rangle$ is better seen in the system that we have considered in this
paper. In the discussion of the multiple gaps and the current peaks
between them, the essential point was that it is possible that,
between two {\it stable insulating} states, there is an {\it unstable
conducting} state that the system has to go through, and the unstable
state is brought upon by the topology of the configuration (two
branches). Whether the geometry of the array is circular or elliptical
should not matter, as long as stable configurations with different
number of trapped electrons are found in the array. One may also note
that the multiple Coulomb gap phenomena is not possible in
one-dimensional arrays with simply-connected geometry. In real
experiments, where the size of the dots in the array should have some
distribution about the average value and the temperature is low yet
finite, we predict that it is mostly likely that the negative
differential conductance regions can be seen in the $I-V$ curve in
place of the additional Coulomb gaps.

% Acknowledgment

This work has been supported by the Ministry of Information and
Communications of Korea.

\end{document}